\documentclass{ws-procs975x65}
\usepackage[utf8]{inputenc}
\usepackage{amsmath, amssymb, amsfonts}

\begin{document}

\title{Eddington inspired Born Infeld Theory: \\A new look to the
matter-coupling paradigm}

\author{T. Delsate${}^{1,2}$, Jan Steinhoff${}^{1}$}
\address{${}^1$ CENTRA, Instituto Superior T\'ecnico, Avenida Rovisco Pais 1,
1049-001 Lisboa, Portugal, EU}
\address{${}^2$ UMons, Universit\'e de Mons, Place du Parc 20, 7000 Mons,
Belgium,
EU}

\bodymatter

\begin{abstract}
We discuss some consequences of changing the matter to gravity coupling
without affecting the gravitational dynamics. The Einstein tensor is usually
assumed to be proportional to the stress tensor due to the divergence free
property of both object. This is not the only consistent way to couple matter to
gravity; we explore some aspect of consistent modification to the matter/gravity
coupling using the recently proposed Eddington inspired Born Infeld extension of
gravity.
\end{abstract}

\section{Motivations}

The theory of General Relativity has raised questions soon after it was
formulated. Even though it passed numerous experimental tests, it still poses
problems at larger scales. For instance, already at the galactic scale, one has
to invoke extra ingredient to understand the galactic rotational curve. On a
more theoretical aspect, Gravity is a non renormalizable theory which is a
major obstacle to a quantum formulation of gravity. All these reasons and many
others call for a completion or modification to General Relativity. There are
essentially three possibilities, either adding new degrees of
freedom, either changing the space-time dynamics or the matter to
gravity coupling. There might be relations between these approaches and they
can be mixed, but we don't discuss this aspect here.
In this proceeding, we focus on the last possibility, namely the matter to
gravity coupling. In particular, we use a recently proposed model to discuss
the coupling, namely the Eddington inspired Born Infeld (EiBI) model
\cite{Banados:Ferreira:2010}. 

\section{EiBI model and the coupling paradigm}
The Eddington inspired Born Infeld model of gravitation
\cite{Banados:Ferreira:2010} is described by the following action
\begin{equation}
S_{EiBI} = \frac{1}{\kappa} \int \left( \sqrt{| g_{ab} + \kappa R_{ab}  |} -
\lambda \sqrt{|g_{ab}|} \right) d^4 x + S_m,
\end{equation}
where $g$ is the metric, $R$ is the Ricci tensor built from an independent
connection, say $S^a_{bc}$, $S_m$ is a minimally coupled matter action,
$\kappa$ is essentially the Newton constant and $\lambda -1$ is the
comsological constant. Variation \'a la Palatini leads to the following set of
equations
\cite{Banados:Ferreira:2010,Vollick:2003qp}
\begin{eqnarray}
 q_{ab} &=& g_{ab} + \kappa R_{ab},\  \sqrt{-q} q^{ab} = \sqrt{-g}\left(
\lambda g^{ab} + \kappa T^{ab}\right) ,\nonumber\\
S^a_{bc} &=& \frac{1}{2}q^{ad}\left( -\partial_d q_{bc} + \partial_b
q_{dc} + \partial_c q_{bd} \right),\ T^{ab} = \frac{1}{\sqrt{-g}}\frac{\delta
S_m}{\delta g_{ab}},
\label{eibieqs}
\end{eqnarray}
where $g$ and $q$ are the determinant of $g_{ab}$ and $q_{ab}$ and where
$q^{ab}, g^{ab}$ are the matrix inverses of $g_{ab},q_{ab}$.
Interestingly, this theory identically reduces to General Relativity in vacuum.
In fact, it does not introduce additional degrees of freedom compared to
General relativity \footnote{it was recently proven \cite{Pani:Sotiriou:2012}
that this formulation of EiBI suffers from specific singular behaviors for
compact objects. In fact, because of the Palatini nature of the model, a
antisymmetric part in the connection should be considered, introducing new
degrees of freedom. In our work we assumed a symmetric connection, yet
introducing no additional degrees of freedom.}. A direct consequence of these
two ingredients is that the theory changes only \emph{inside} matter.
This leads to a bunch of interesting consequences, essentially regarding
singularities in General Relativity. For example, it was shown
\cite{Banados:Ferreira:2010} that the Big-Bang singularity might be avoided
with such a model. It was also argued that astrophysical singularities may be
avoided or softened in black hole formation processes
\cite{Pani:Cardoso:Delsate:2011,Pani:Delsate:Cardoso:2012}

\subsection{Reformulation of EiBI}
The EiBi model of gravity can be reformulated in a way that makes the
modification with respect to General Relativity very explicit. In particular,
the case of perfect fluids matter field is quite enlightening
\cite{Delsate:Steinhoff:2012}.
Indeed, equations \eqref{eibieqs} can be rewritten in the following form
\begin{equation}
G^a{}_b[q_{cd}] = R^a{}_b - \frac{1}{2} \delta^a{}_b R = \gamma\mathcal{T}^a{}_b
- \Lambda\delta^a{}_b ,
\label{einstein_q}
\end{equation}
where $\tau = \sqrt{g/q}$ and where we defined the apparent stress tensor
$\mathcal{T}^a{}_b$ defined as
\begin{equation}
\mathcal{T}^a{}_b = \tau T^a{}_b + \mathcal{P} \delta^a{}_b , \
\gamma\kappa \mathcal{P} = \tau - 1 - \frac{1}{2} \tau \gamma\kappa T ,
\end{equation}
and the cosmological constant $\Lambda = (\lambda - 1) / \kappa$.
$\mathcal{P}$ plays the role of an isotropic pressure addition.
For the case of a perfect fluid, the metric $g$ can be eliminated explicitly,
leaving a theory of pure General relativity coupled to a perfect fluid with a
modified equation of state. More precisely, the metric determinant ratio is
given by $ \tau = \left[\mbox{det}\left( \delta^a{}_b -\kappa \gamma T^a{}_b
\right)\right]^{-\frac{1}{2}}$.

This reformulation shows explicitly, at least in the case of a perfect fluid
how the coupling between gravity and matter is changed in this theory. 
It should be noted however that in this reformulated case, matter is coupled
(in a modified way)  to the metric $q$, which is an auxiliary metric in the
original theory. However, since observations are usually performed from outside
matter, since both metric are the same in vacuum, the effect of the modified
coupling to $q$ affects measurements performed with $g$ in vacuum indirectly.
Interestingly enough, all the nice features of the EiBI model can be understood
in terms of the reformulated version and of the modification in the coupling
\cite{Delsate:Steinhoff:2012}.
This motivates further studies in 'pure' coupling modification.
It should be mentioned that the set of equivalent equations \eqref{einstein_q}
can be obtained from the following action principle:
\begin{equation}
 S_{equiv} =\frac{1}{\gamma}\int \! d^4x \sqrt{-q} \left[ R[q]  - 2
\frac{\lambda}{\kappa} + \frac{1}{\kappa} ( q^{ab}g_{ab} - 2 \tau ) \right] \! 
 + S_M(g),
\label{equiv_action}
\end{equation}
where $S_M(g)$ is minimally coupled to $g$. This action is very similar to
a bi-metric theory, apart from the fact that the metric $g$ is not dynamical.

This is the particular case of the EiBI theory of gravity, but it is tempting
to use this action as a basis for investigating coupling modifications.

\section{Conclusion}
Theories effectively changing the coupling between matter and gravity pass most
experimental tests of General Relativity, provided they are equivalent in
vacuum. Note that most of the tests of Gravity are in fact testing General
Relativity in vacuum. An interesting realization of such a theory is the
Eddington Born Infeld theory of gravity. We made an explicit correspondence
between the modification of the matter / Gravity coupling and the interesting
features of the EiBI model. In particular we showed that in the case of a
perfect fluid, it is possible to fully reexpress the theory as a pure General
Relativistic model with a modification to the equation of state of the fluid. As
a consequence, the fluid, keeps all its standard properties, except from a
gravitational point of view where the geometry sees it differently. This results
in an 'apparent' matter content.

It should be mentioned that the resulting apparent source displays interesting
features towards energy conditions (see detailed discussion in
\cite{Delsate:Steinhoff:2012}).\\

This work was supported by DFG (Germany) through project STE 2017/1-1,
FCT (Portugal) through projects PTDC/CTEAST/098034/2008 and PTDC/FIS/098032/2008
and CERN through project CERN/FP/123593/2011.

\end{document}